# Numerical Modelling of Buffer Layers for Advancing CZTSSe Solar Cell Efficiency

Tanzir Ahamed[1,2,*], Fozlur Rayhan[3], Imteaz Rahaman[4], Md. Hamidur Rahman[3], Md. Mehedi Hasan Bappy[1,2], Tanvir Ahammed[5], Sampad Ghosh[2]

[1]Department of Electrical and Electronic Engineering, CCN University of Science and Technology, Cumilla-3503, Bangladesh

[2]Department of Electrical and Electronic Engineering, Chittagong University of Engineering and Technology, Chattogram-4349, Bangladesh

[3]Department of Electrical and Computer Engineering, Lamar University, Beaumont, TX, 77705, USA

[4]Department of Electrical and Computer Engineering, University of Utah, Salt Lake City, UT, 84112, USA

[5]Department of Materials Science and Engineering, University of Rajshahi, Rajshahi-6205, Bangladesh

**\*Corresponding author:** *Tanzir Ahamed (tanzir.eee2k15@gmail.com)*

***Abstract:*** Kesterite is a leading candidate among inorganic thin-film photovoltaic technologies, offering sustainable and environmentally friendly solutions without reliance on critical raw materials. This study investigates the performance of CZTSSe-based kesterite solar cells using SCAPS-1D simulations. Four device configurations are analyzed by integrating the CZTSSe absorber layer with buffer materials, including CdS, $SnS_2$, IGZO, and ZnSe, selected based on their energy band alignment. Moreover, key parameters influencing device efficiency, such as absorber defect density, buffer layer thickness, and the donor and defect densities of the buffer materials, are systematically examined. The thickness of the absorber layer and acceptor density are optimized, considering practical manufacturing constraints. Following optimization, the i-ZnO/$SnS_2$/CZTSSe/Au configuration achieves a notable efficiency of 28.38%, with a $V_{OC}$ of 0.83 V, a $J_{SC}$ of 39.93 mA/cm$^2$, and a fill factor of 85.4%. Furthermore, the stability of the optimized structures is evaluated under varying conditions, including resistances, temperature, generation and recombination dynamics, as well as JV and QE characteristics. These findings provide valuable insights for advancing the efficiency and stability of CZTSSe solar cells, contributing to the development of sustainable photovoltaic technologies.

## 1. Introduction:

Solar energy is a sustainable and eco-friendly alternative to conventional natural resources, providing a viable answer to global energy demands. Solar cell technology has become an increasingly popular and established technique for electrical energy conversion among renewable energy sources in recent decades. [1]. Silicon-based solar cells, traditionally the most widely used and efficient in the photovoltaic (PV) industry [2], face significant cost challenges in their fabrication. Consequently, researchers are shifting their focus to second and third-generation thin-film solar cell technologies, such as copper indium gallium selenide (CIGS) [3], [4], amorphous silicon (a-Si) [5], perovskite [6], [7], cadmium telluride (CdTe) [8] and kesterite based copper zinc tin sulphide (CZTS) [9], [10] etc. These alternatives provide feasible solutions and expected upgrades over silicon-based solar cells, with the objectives of producing low-cost, high-efficiency, and durable solar cells. To predict the expected outcomes for serving those essential objectives, theoretical methodologies using developed software packages and advanced computing with machine learning are devised to evaluate the anticipated performance of every specifically designed cell [11], [12]. In the contemporary arena, CZTS solar cells have garnered significant attention in thin-film technology due to their numerous advantages, including exceptional optoelectronic properties and a broad light absorption spectrum. With an absorption coefficient exceeding $10^4$ cm$^{-1}$ [13], they have been widely utilized by the scientific community across various fields for decades. This interest has driven the development of high-quality absorber materials such as copper zinc tin sulfoselenide (CZTSSe), which features a tunable bandgap ranging from 1 to 1.5 eV [14]. Additionally, a CZTSSe absorber layer with a thickness of just 1–2 µm can achieve excellent power conversion efficiency [15]. The scientific community considers CZTSSe thin-film solar cells as a potential replacement for existing absorber layers such as CIGSSe and CdTe. Moreover, CZTSSe is composed of non-toxic and cost-effective materials, making it an environmentally sustainable option. CZTSSe exhibits two crystalline structures: stannite and kesterite, with the kesterite structure being more stable and reliable [16]. It can be synthesized using a variety of techniques, including electrodeposition, spray pyrolysis [17], sputtering [18], atomic layer deposition [19], and spin coating [20], among others. However, optimizing device design is essential to achieving high conversion efficiency, enabling the development of highly efficient solar cells for future applications [21], [22]. A recent study by Pakštas *et al.* [23] reported that incorporating an $Sb_2Se_3$ buffer layer and a low-temperature selenium process improved the efficiency of CZTSSe-based solar cells by up to 3.48%. Similarly, CZTSe solar cells achieved a 12.6% efficiency using a high vapor transport deposition technique, which combines precursor material deposition with sulfur-selenization in the same chamber [24], [25]. This approach is notable for its rapid deposition rate and precise chemical composition control, contributing to the observed efficiency. In another recent work, Ahmad *et al.* [26] demonstrated that atomic layer deposition of a zinc-tin oxide (ZTO) buffer layer in silver-doped

CZTSSe solar cells resulted in a power conversion efficiency (PCE) of 11.8% [26]. This represents one of the highest efficiencies reported for Cd-free kesterite thin-film solar cells to date. Additionally, theoretical studies have predicted significant efficiencies for CZTSSe solar cells with optimized absorber bandgaps, underscoring the material's potential for high-performance applications.

The buffer layer plays a critical role in enhancing the performance of thin-film solar cells. It facilitates proper band alignment between the CZTSSe absorber and the transparent conducting oxide (TCO), enabling efficient electron transport while reducing recombination losses [27], [28]. Additionally, the buffer layer passivates surface defects on the absorber layer, thereby decreasing carrier recombination and improving both the open-circuit voltage ($V_{OC}$) and the fill factor (FF) [29]. Its high optical transparency ensures that most photons reach the absorber layer, enhancing light absorption and boosting the cell's overall power output. Moreover, the buffer layer acts as an insulating barrier, preventing shunt pathways and protecting against efficiency loss [30]. By mitigating energy barriers and smoothing the absorber-TCO interface, the buffer layer ensures consistent charge separation and extraction [31]. As such, a suitable buffer layer is essential for achieving high-efficiency, stable, and reliable CZTSSe solar cells. Continuous advancements over recent years have demonstrated significant contributions of buffer materials in thin-film solar cells, including CZTSSe and CdTe. These contributions include proper band alignment, improved interface quality, enhanced optical absorption, and performance optimization. Additionally, the buffer layer reduces defects and interfacial strain associated with the window layer, as has been well established [32]. Studies reveal that omitting a buffer layer during fabrication significantly decreases efficiency, highlighting the critical impact of its energy bandgap and thickness on the overall performance of solar cells [33], [34]. Notable advancements in buffer layer design include a promising Al/TCO/IGZO/$C_2$N/CZT/Pt structure reported by Taseen *et al.* [35], achieving an efficiency of 28.16% through the use of IGZO as a buffer layer. The excellent stability of IGZO across various optical applications further underscores its utility [36], [37]. Et-Taya *et al.* [38] demonstrated the compatibility of ZnSe, $TiO_2$, and CdS buffer layers with CZTSSe absorber, achieving simulated efficiencies of 22.42%, 23.13%, and 23.16%, respectively. Additionally, theoretical studies have shown that buffer materials like CZGSe, $CuSbS_2$, and $Sb_2S_3$ can achieve efficiencies of 30.52%, 31.21%, and 31.29%, respectively, approaching the Shockley-Queisser (SQ) limit [39]. Tripathi *et al.* [40] highlighted the potential of hybrid buffer layers to exceed the performance of conventional single-layer materials like CdS. Their optimization achieved a simulated efficiency of 16.52% using a hybrid buffer layer of CdS and $In_2S_3$, with both components having a thickness of 40 nm. These findings demonstrate that more optimized and efficient buffer materials are available, offering significant potential for improving the power conversion efficiency (PCE) of practical solar cells.

In this study, we focus on the optimization and performance analysis of four different buffer layers in CZTSSe solar cells. The effects of thickness variation, defect density, series and shunt resistance, and temperature on performance parameters are comprehensively examined. Additionally, the J-V curve and quantum efficiency (QE) are analysed to evaluate the electrical and optical performance of the simulated cell configurations. Finally, the generation-recombination profile is investigated to understand carrier dynamics, efficiency optimization, and potential performance limitations of the cell.

## 2. Materials and methodology

### 2.1. CZTSSe-Based Solar Cell Structure

The proposed structure comprises four fundamental layers. Fig. 1(a) illustrates the schematic representation of the simulated structure, where CZTSSe serves as the primary absorber layer.

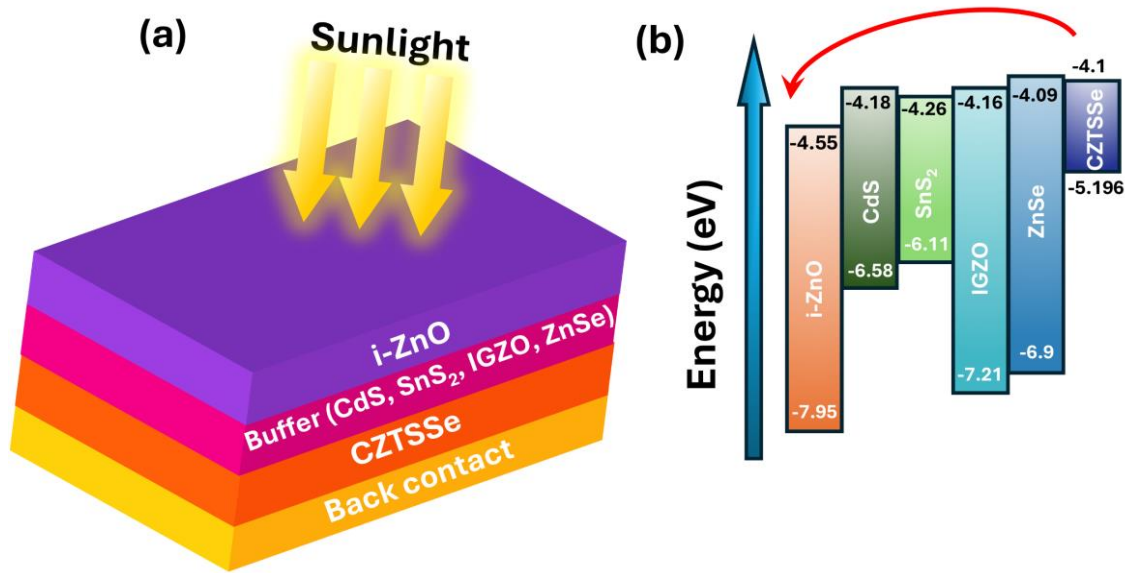

**Fig. 1.** (a) Schematic of the simulated CZTSSe solar cell structure, showing key layers. (b) Energy band alignment of the layers, illustrating conduction band offsets for efficient charge transport.

Four distinct buffer materials—CdS, SnS$_2$, IGZO, and ZnSe—are identified as effective carrier transporters based on prior research. Sunlight initially interacts with the i-ZnO layer on the glass substrate, followed by the buffer layer, which facilitates photon transmission. The absorber layer captures a significant portion of the incident light. For this simulation, standard solar irradiation (AM 1.5G) is considered, providing an incident power of 1000 W/m². The initial simulation temperature is set at 300 K. Fig. 1(b) shows the energy band alignment of the proposed solar cells. In an optimal solar cell configuration, the HOMO of the buffer material must be more negative than that of the absorber material, as depicted in Fig. 1(b). The conduction band offsets (CBO) for CdS, SnS$_2$, IGZO, and ZnSe are 0.08 eV, 0.16 eV, 0.06 eV, and -0.01 eV, respectively, facilitating efficient electron transport across the structure. Optimal band alignment is critical for enhancing cell efficiency. Within this structure, the i-ZnO/CZTSSe/SnS$_2$/back contact configuration exhibits the most favourable and appropriate band alignment, contributing to improved device performance.

Table.1 represents the numerical values for the simulated structure. All the adopted numerical values are collected from previously published literatures. Those values are the key ingredients for the simulation for SCAPS-1D. Supplying those data into the SCAPS-1D program leads to the performance evaluation of the structures.

**Table 1.** Data table for initial electrical properties of all layers.

| Parameters | CZTSSe[41] | SnS$_2$[42] | IGZO[43] | CdS[44] | ZnSe[45] | i-ZnO |
|---|---|---|---|---|---|---|
| Thickness(nm) | 1000 | 150 | 70 | 50 | 70 | 80 |
| Band Gap (eV) | 1.096 | 1.85 | 3.05 | 2.4 | 2.81 | 3.4 |
| Electron affinity (eV) | 4.1 | 4.26 | 4.16 | 4.18 | 4.09 | 4.55 |
| Dielectric permittivity (relative) | 13.6 | 17.7 | 10 | 10 | 8.6 | 10 |
| CB effective density of states (cm$^{-3}$) | 2.2×10$^{18}$ | 7.32×10$^{18}$ | 5×10$^{18}$ | 2.2×10$^{18}$ | 2.2×10$^{18}$ | 4×10$^{18}$ |
| VB effective density of states (cm$^{-3}$) | 1.8×10$^{19}$ | 1×10$^{19}$ | 5×10$^{18}$ | 1.9×10$^{19}$ | 1.8×10$^{18}$ | 9×10$^{18}$ |
| Electron thermal velocity (cm/s) | 1×10$^7$ | 1×10$^7$ | 1×10$^7$ | 1×10$^7$ | 1×10$^7$ | 1×10$^7$ |
| Hole thermal velocity (cm/s) | 1×10$^7$ | 1×10$^7$ | 1×10$^7$ | 1×10$^7$ | 1×10$^7$ | 1×10$^7$ |
| Electron mobility (cm$^2$/Vs) | 100 | 50 | 15 | 100 | 4×10$^2$ | 50 |
| Hole mobility (cm$^2$/Vs) | 25 | 25 | 0.1 | 25 | 1.1×10$^2$ | 20 |
| Shallow uniform donor density, N$_D$ (cm$^{-3}$) | 0 | 9.85×10$^{19}$ | 1×10$^{18}$ | 1×10$^{18}$ | 1×10$^{18}$ | 5×10$^{17}$ |
| Shallow uniform acceptor density, N$_A$ (cm$^{-3}$) | 1×10$^{15}$ | 0 | 0 | 0 | 0 | 0 |
| Defect Nt (1/cm$^3$) | S(D): 1×10$^{15}$ | S(D): 1×10$^{14}$ | S(D): 1×10$^{14}$ | S(D): 1×10$^{15}$ | S(D): 1×10$^{15}$ | S(D): 1×10$^{15}$ |

Here, S(D)=single donor

## 2.2. Methodology

SCAPS-1D is a vital tool for analyzing and improving the characteristics of thin-film solar cells, serving as a valuable resource for advancing solar cell technology. It solves key equations in the background, including Poisson's equation, steady-state electron-hole continuity equations, and the current density equations for electrons and holes. These equations are represented as Equations (1) through (5).

$$\frac{d^2\psi(x)}{dx^2} + \frac{q}{\varepsilon_0\varepsilon_r}[\mathcal{P}(x) - n(x) + N_D^+(x) - N_D^-(x) + \mathcal{P}_t(x) - n_t(x)] = 0 \quad (1)$$

Where potential electrostatic potential is represented by ψ, ionized donor concentration by $N_D^+$, ionized density acceptor by $N_D^-$, electron and hole densities by $n$ and $p$, relative permittivity by $\varepsilon_0$ and vacuum permittivity by $\varepsilon_r$, trapped electron and hole representation by $p_t$ and $n_t$, electron charge by q, and position in the x-coordinate by $x$.

$$-\frac{dJ_n}{dx} + G - R = 0 \quad (2)$$

$$-\frac{dJ_p}{dx} + G - R = 0 \quad (3)$$

$$J_n = qn\mu_n E + qD_n \frac{\partial n}{\partial x} \quad (4)$$

$$J_p = qn\mu_p E + qD_p \frac{\partial p}{\partial x} \quad (5)$$

Here, G = carrier generation rate, R = net recombination from the direct and indirect band, $J_n$ and $J_p$ = current densities of hole and electron, $D_n$ and $D_p$ both are electron diffusion coefficients for hole and electrons, E = electric field, $\mu_p$ and $\mu_n$ = both are mobilities of hole and electron accordingly.

The simulation settings and input numerical parameters, including bandgap, electron affinity, dielectric permittivity, electron mobility, and hole mobility, are established for each layer of the solar cell. These values are derived from prior research. Once these parameters are provided, performance indicators such as current voltage (I-V), capacitance voltage (C-V), and quantum efficiency (QE) are assessed to evaluate the electrical and optical performance of the proposed cell.

## 2.3. Energy band diagram of CZTSSe based solar cells

The band diagram provides a clear visualization of the electrical characteristics, material performance of individual layers, and overall device efficiency. Optimal band alignment facilitates efficient carrier transport, minimizes recombination losses, and enhances overall power conversion efficiency (PCE). Fig. 2 illustrates various alignments with all potential layers, including four distinct buffer layers. The proposed architecture comprises a glass substrate, a buffer layer (CdS, SnS$_2$, IGZO, or ZnSe), an absorber layer (CZTSSe), and a back contact. All layers exhibit suitable band alignment. Fig. 2(b) highlights the highest band offset of 0.16 eV relative to the primary absorber, CZTSSe. Two types of band offsets are identified in this study: the positive offset, often referred to as a spike type, and the negative offset, known as a cliff type. The spike-type band offset at the interface between the buffer and absorber layers enhances interface stability, promotes carrier transport, and reduces leakage current. Furthermore, this type of offset enables band structure engineering to improve device performance and may induce hot electron effects in ultrafast optoelectronic devices.

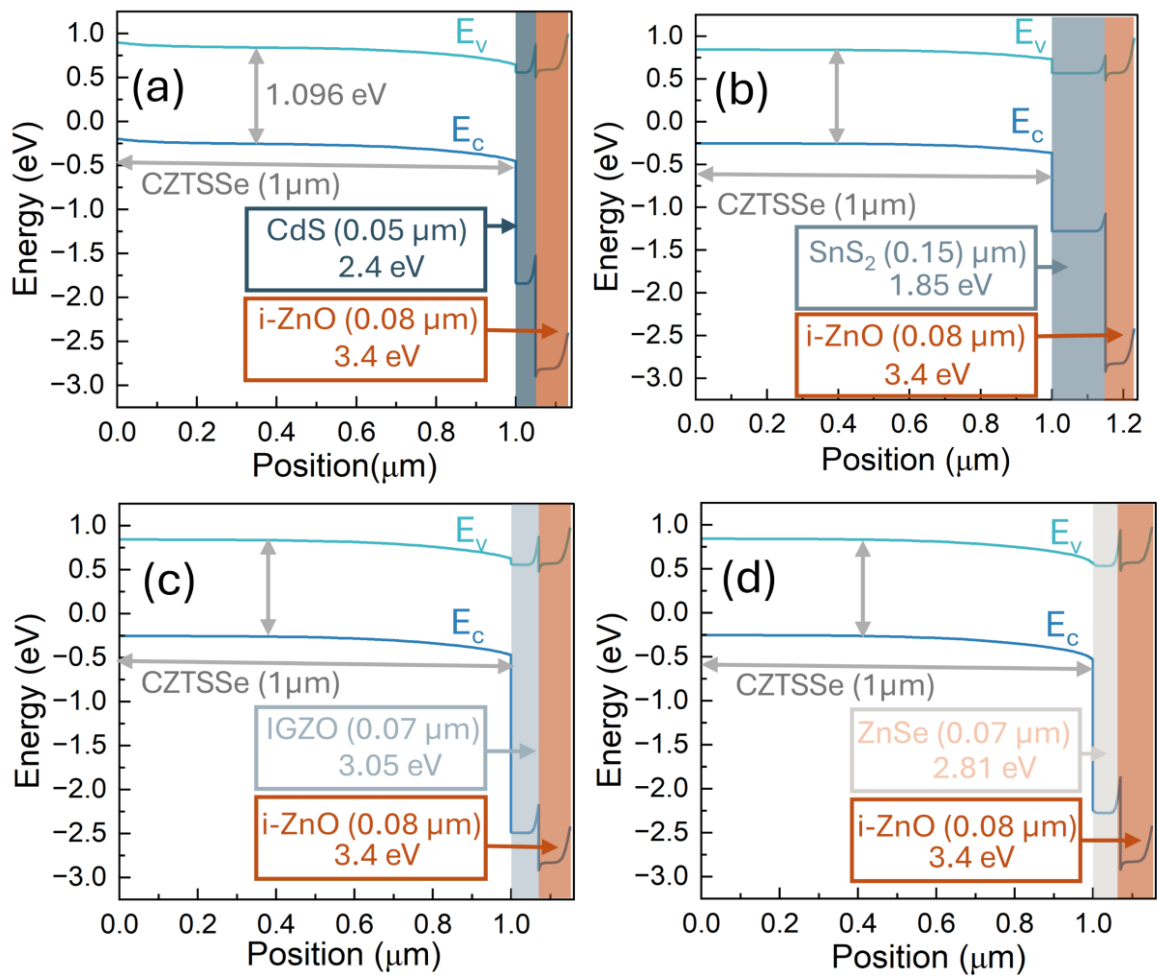

**Fig. 2.** Energy band diagrams for CZTSSe solar cells with different buffer materials: (a) CdS, (b) SnS$_2$, (c) IGZO, and (d) ZnSe. Each diagram illustrates the energy levels of the conduction

band ($E_C$) and valence band ($E_V$) across the device structure, highlighting the band alignment between the i-ZnO, buffer, CZTSSe absorber, and back contact layers.

## 3. Result and discussion

### 3.1. Effects of defect density of Absorber and buffer layer thickness

The effectiveness of all structures appears to be mostly independent of the buffer layer thickness. With an increase in defect density of the absorber, the power conversion efficiency (PCE) rises with an increase in $N_t$ at $10^{12}$ cm$^{-3}$, the power conversion efficiencies (PCEs) were around 9%, and at $10^{16}$ cm$^{-3}$, the value reached an exceptional level of almost 19%. This section is excluded from the optimization segment since the greatest efficiency does not surpass the initial efficiency. Although the thickness of the buffer layers hardly affects the PCE, it is advisable to maintain the value as low as possible due to the cost of manufacturing.

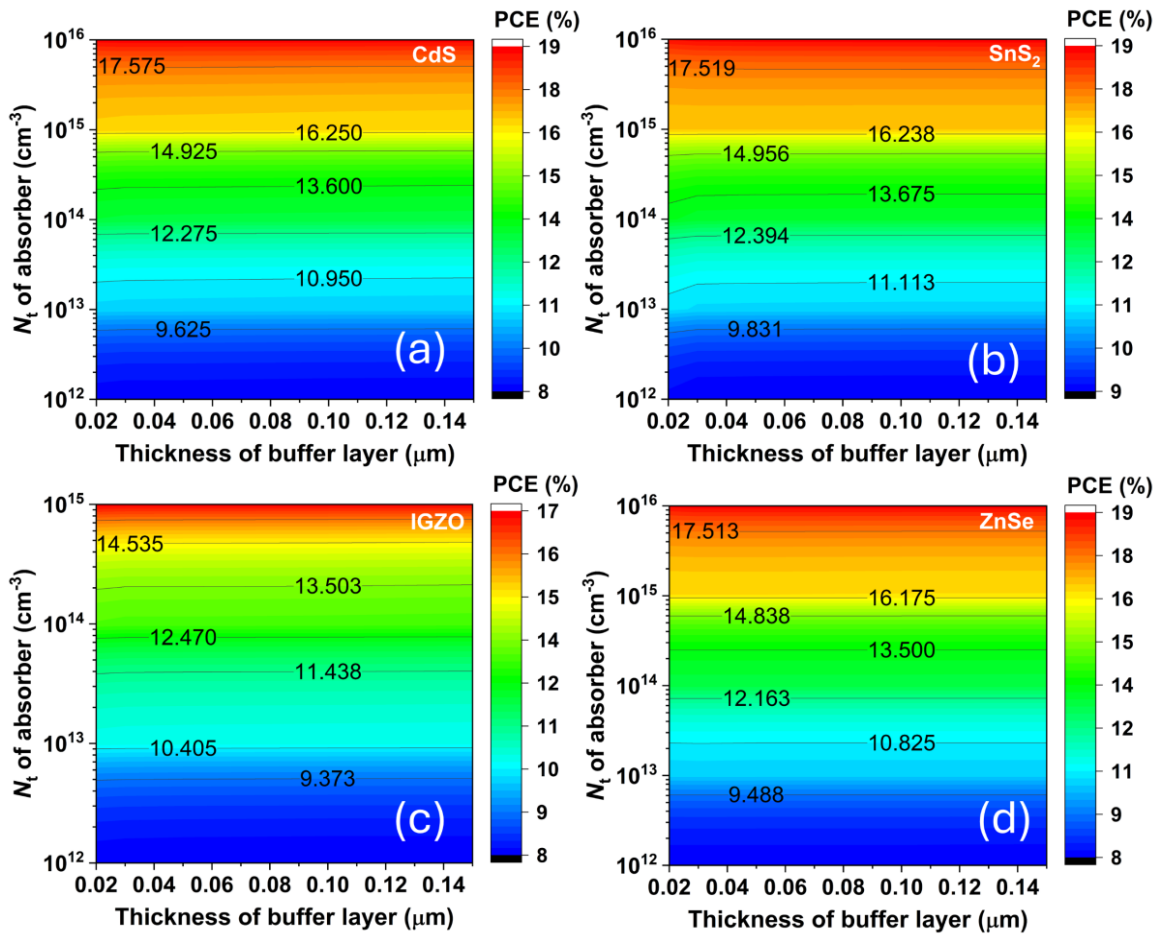

**Fig. 3.** Contour graphs for the distinct buffer materials (a) CdS (b) SnS$_2$ (c) IGZO (d) ZnSe for the influence of buffer thickness and absorber defect density variation on PCE.

Fig.3(a-d) illustrates that the buffer thickness has been altered from 20 nm to 150 nm, while the defect density $N_t$ has been altered from $10^{12}$ to $10^{16}$ cm$^{-3}$. Hong *et al.* stated in [46] that a buffer thickness of 70 nm resulted in an efficiency of 8.77% using a CZTSSe (1.16 eV) absorber in an experimental cell. Similarly, Fig. 3 demonstrates that buffer thickness has a negligible impact on the PCE of the cell. This may seem to have the same effect on all four buffer materials. The density of absorber defects has a significant and substantial effect on performance. The maximum power conversion efficiency (PCE) reported is 18.88% for CdS, with a buffer layer thickness of 30 nm and a defect density ($N_t$) of $10^{16}$ cm$^{-3}$.

### 3.2. Effect of buffers' donor and defect density

In general, higher donor density of the buffer layer helps to generate the carrier in the material, and higher defect density may cause recombination rates among the carriers, resulting in decreasing free carriers. Fig. 4(a)-(d) illustrates the variation of PCE with the change in donor and defect density of the buffer layer. The values of $N_D$ and $N_t$ are varied from 1016 cm-3 to $10^{20}$ cm$^{-3}$ and $10^{12}$ cm$^{-3}$ to $10^{16}$ cm$^{-3}$, respectively. For the four structures, the PCEs show a higher value at ND of >1017 cm-3, and the efficiencies do not depend on the defect density of buffer layers. However, when SnS$_2$ is used as buffer material, it seems to be a little dependent on the $N_t$ of the material. At $N_t$ of < $10^{15}$ cm$^{-3}$, the structure exhibits comparatively higher PCE. Fig. 4 highlights the variation of PCE depending on buffer donor density ($N_D$) and buffer defect density ($N_t$). The variation taken for the ND is 1016 to 1018 cm-3, and $N_t$ is from $10^{16}$ to $10^{20}$ cm$^{-3}$. The analysis reveals that both the effect of $N_D$ and $N_t$ effect on PCE have been examined. Overall, the effect of both donor and defect densities on PCE is limited for the proposed structure, with minor variations depending on the specific buffer material.

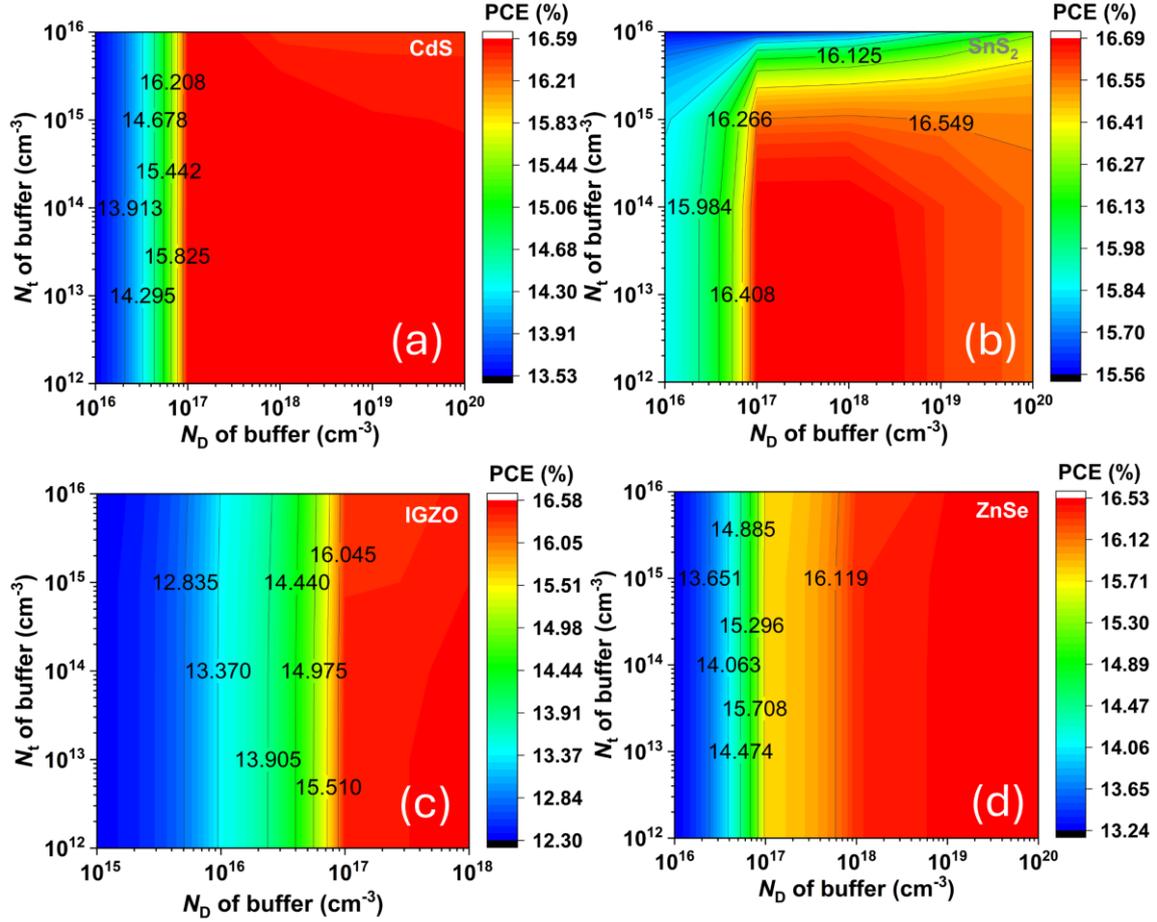

**Fig. 4.** Effect of buffer donor density ($N_D$) and defect density ($N_t$) on PCE for simulated buffer materials: (a) CdS, (b) $SnS_2$, (c) IGZO, and (d) ZnSe. PCE improves at $N_D>10^{17}$ cm$^{-3}$, with minimal impact from $N_t$, except for $SnS_2$ at lower defect densities.

### 3.3. Optimization of absorber layer thickness

The thickness of the absorber is critical for enhancing the performance features of a solar cell. The simulation was conducted in order to evaluate the effects of absorber thickness and to optimize layer thickness. The increase in absorber thickness results in a greater number of photons reaching the absorber, hence significantly enhancing the PCE. Conversely, a thicker absorber results in increased recombination loss, elevated defect density, interface optical loss, and higher manufacturing expenses [46]. Consequently, optimization is essential for reducing manufacturing costs while enhancing power conversion efficiency (PCE) in practical cells. Srivastava *et al.* [47] studied that the highest amount of efficiency was found at 19.59%, with CZTSe as the absorber thickness of 1000nm. This is the primary reference to tune the thickness around 800nm to 1600nm for our study. All structures of the various buffer materials exhibit a rise in all performance, indicated by an increase in absorber thickness. For this simulation, the optimal cell performance among the four structures is noted as 1600nm with a PCE of 17.69% for SnS$_2$. The optimal performance for all materials occurs at a thickness of 1600 nm, with measured PCE of 17.68%, 17.67%, and 17.53% for CdS, IGZO, and ZnSe, respectively. Beyond the optimized thickness, efficiency improvement is not significant, and a thicker absorber can lead to a greater manufacturing cost and less performance due to many losses. The rest of the performance parameters are found with SnS$_2$ of 0.528V, 43.29 mA/cm$^2$, and 77.37% for $V_{OC}$, $J_{SC}$, and FF, respectively.

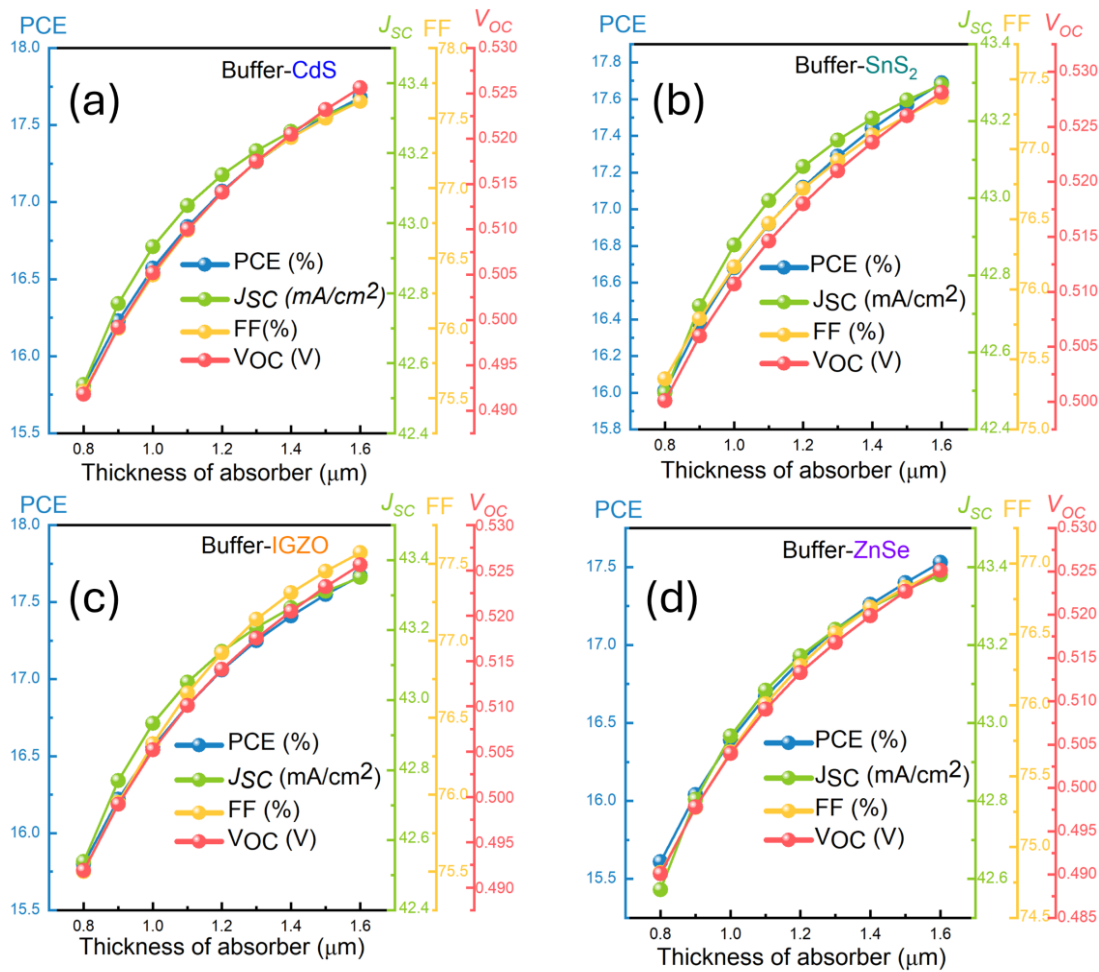

**Fig. 5.** Effect of absorber thickness variation on the performance of solar cell configurations with different buffer materials: (a) CdS, (b) SnS$_2$, (c) IGZO, and (d) ZnSe.

### 3.4. Optimization of acceptor density for absorber layer

Fig.6 illustrates the impact of varying the acceptor density of the absorber on the performance parameters of solar cells for four different buffer materials: (a) CdS, (b) SnS$_2$, (c) IGZO, and (d) ZnSe. The power conversion efficiency (PCE) increases significantly with higher acceptor density, starting at approximately 17.5% at and reaching a maximum of about

28% near for all configurations. The short-circuit current density ($J_{SC}$) remains stable, varying between 39 mA/cm² to 44 mA/cm² at the simulation range from $10^{15}$ cm$^{-3}$ to $10^{20}$ cm$^{-3}$. Among of the structure, the configuration with buffer ZnSe shows the highest amount of $J_{SC}$, 43.38 mA/cm² at $10^{15}$ cm$^{-3}$. The fill factor (FF) shows a slight decline, starting from the highest value at approximately 85.4% with buffer SnS2 configuration at an absorber density of $10^{20}$ cm$^{-3}$ and decreasing to around 76.9% with ZnSe at $10^{15}$ cm$^{-3}$. The open-circuit voltage ($V_{OC}$) demonstrates a steady increase, starting from 0.5 V at $10^{15}$ cm$^{-3}$, reaching near 0.83 V at $10^{20}$ cm$^{-3}$, and stabilizing thereafter. These trends highlight the critical role of absorber acceptor density in influencing solar cell performance, with optimal and best results observed at $10^{20}$ cm$^{-3}$. Subtle differences are observed across the buffer materials, particularly in the rate of improvement for PCE, with SnS$_2$ showing slightly higher PCE under certain conditions.

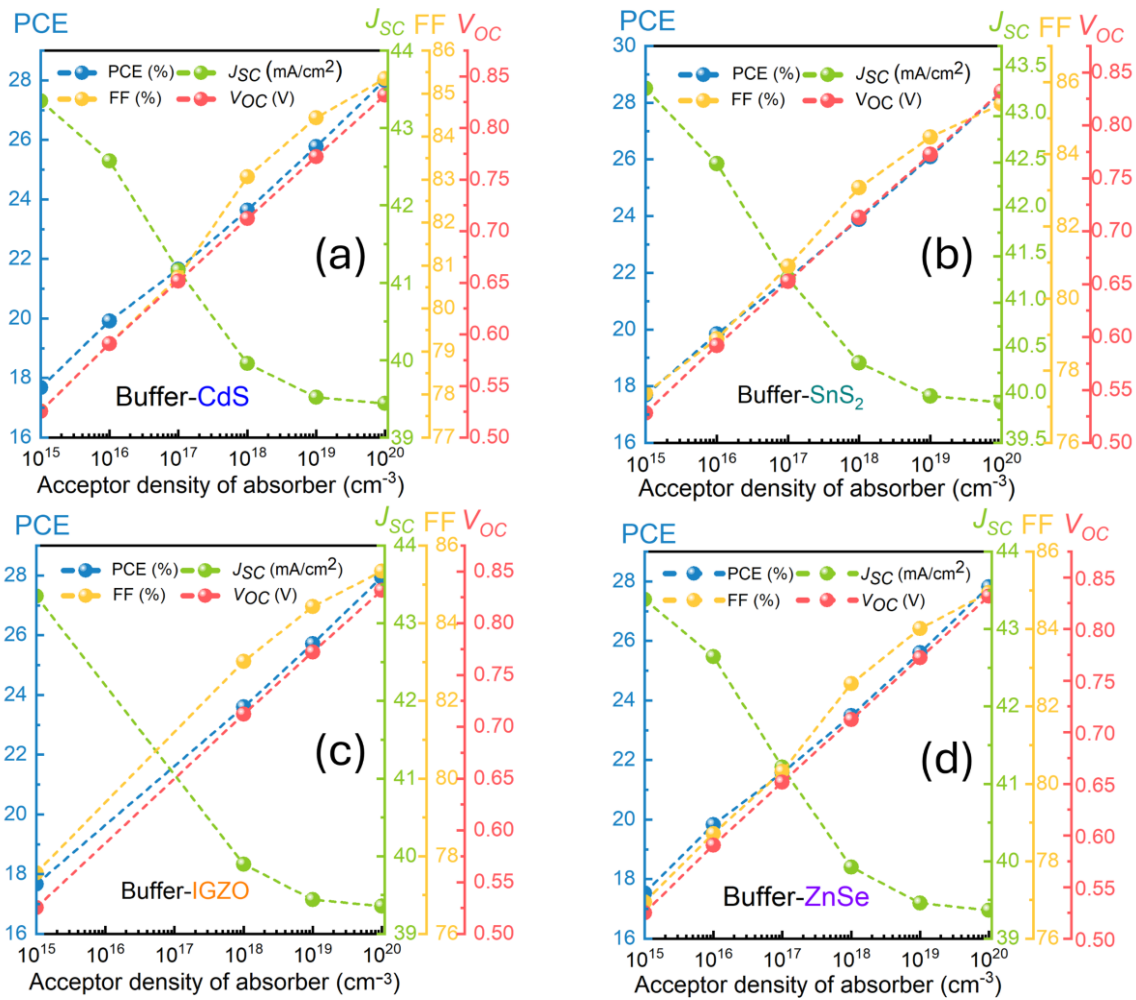

**Fig. 6.** Impact of performance parameters variation depending on changes in acceptor density of the absorber across all four configurations: (a) CdS (b) SnS$_2$ (c) IGZO (d) ZnSe.

### 3.5. Effect of series and shunt resistance

The influence of shunt and series resistance is extremely important for improving the efficiency of cell-building processes. To optimize efficiency and fill factor, the series resistance must be minimized while the shunt resistance should be maximized. Increasing the shunt resistance to its maximum simulated value while minimizing the series resistance in the structure greatly influences the power conversion efficiency, resulting in optimal performance [48]. The simulated range for 1 to 6 Ω-cm² for the series resistance and 10 to $10^6$ Ω-cm² for the shunt resistance in this study. Theoretically, the series resistance must be zero for maintaining the fill factor 100%. The mathematical formula for the understanding the impact of series and shunt resistance can be expressed by the following formula in (6)

$$I = I_{ph} - I_o \left[ exp\left(\frac{V+IR_s}{nV_t}\right) - 1 \right] - \left(\frac{V+IR_s}{R_{sh}}\right) \tag{6}$$

Where I, V, $V_t$, $R_s$, and $R_{sh}$ denote the current, voltage, thermal voltage, series resistance, and shunt resistance, respectively. $I_0$ represents the saturation current, $I_{ph}$ denotes the photo-generated current, and '*n*' represents the ideality factor.

In this part of this study, the impact of shunt and series resistance has been examined though altering the shunt and series resistance. The effect of shunt and series resistance is illustrated in Fig.7 for the simulated structure. An increased leakage current through shunt resistance which is eventually called the shunt current can generate hotspot and heating in the module[49]. For each buffer material, the PCE is significantly impacted when the series resistance is increased. As a consequence, the PCE is declined rapidly. The reverse impact can be observed for the shunt resistance. For an ideal cell, the shunt resistance should be infinite. Previously, Mazumder *et al.* in [50] reported that an amount of 20.34% PCE was found by the tuning of series and shunt resistance of 1.97 Ω-cm² and 1424.05 Ω-cm² accordingly for the AZO/i-ZnO/ZnS/CZTS-bilayer solar structure. This work indicates the primary simulation boundary of our proposed cell. The optimal efficiency for our work has been determined for the structure displaying a minimum series resistance of 1 Ω-cm², validating the mathematical calculations. The series resistance of the primary simulated structure is regarded as 1 Ω-cm² for comparison with the actual cell. Among buffer materials, $SnS_2$ has achieved the maximum power conversion efficiency (PCE) of 26.9%, boasting an optimal simulated shunt resistance and lowest series resistance.

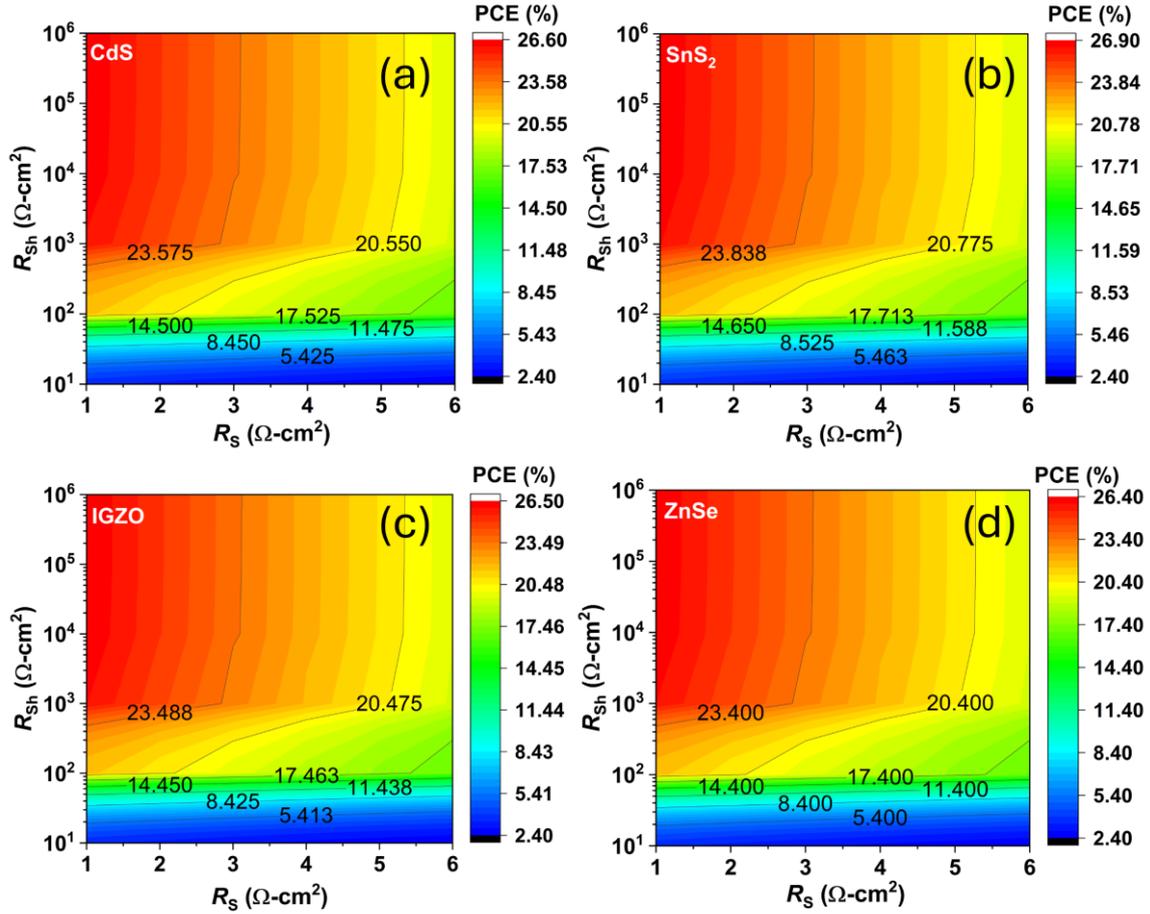

**Fig. 7.** Impact of parasitic resistances (series & shunt) on the PCE for all individual structure.

The optimal efficiency is achieved with the minimum simulated series resistance and maximum shunt resistance of $10^6$ Ω-cm², providing numerical values of 26.6%, 26.9%, 26.5%, and 26.38% for CdS, SnS$_2$, IGZO, and ZnSe, respectively.

### 3.6. Effect of temperature

Fig. 8 illustrates the effects on several performance parameters, including $V_{OC}$, $J_{SC}$, FF, and PCE, for four buffer materials coupled with CZTSSe as the primary absorber. Temperature has been recognized as a crucial element affecting the fluctuation of PCE. The starting temperature for this study is set at 300K. Moreover, the effect of temperature fluctuation may be shown across a range of 300K to 450K. Apart from the performance parameters, current density $J_{SC}$, the rest of the performance parameters, including VOC, FF, and PCE, declined with respect to increasing cell temperature drastically. The main factor for the decrement of other performance parameters is the high dependency on bandgap vs. temperature characteristics. A decrease in the band gap of the semiconductor material is the result of an increase in temperature[4].

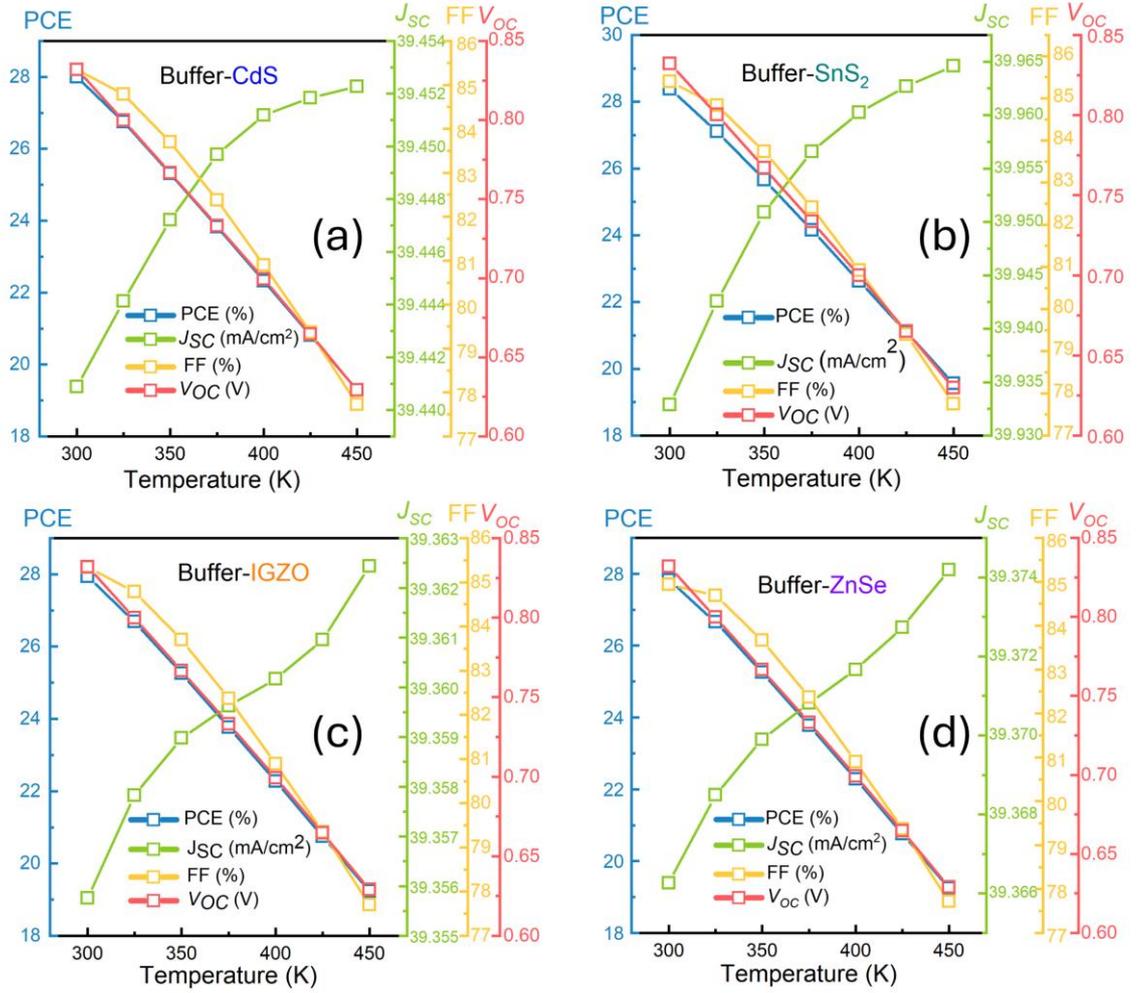

**Fig. 8.** Effect of temperature on the performance of solar cells with different buffer layer configurations: (a) CdS, (b) SnS$_2$, (c) IGZO, and (d) ZnSe. The analysis highlights the influence of temperature variations on the PCE and other key performance parameters for each buffer material.

The current density $J_{SC}$ is slightly increased for the increasing absorption of photons of lower energy while increasing temperature. This reduction in energy reduces the ability of captured photons to produce free charge carriers, thereby minimizing the $V_{OC}$ of the cell. Furthermore, the fill factor is reduced, and the series resistance of the cell is increased as a result of a drop in carrier mobility with temperature. The reduction in $V_{OC}$ and FF results in a decrease in the solar cell's overall efficiency. Consequently, it can be concluded that solar cells perform at their optimum at ambient temperature.

The mathematical expression for the open circuit voltage can be expressed as (7)

$$V_{OC} = \frac{KT}{q}\ln(\frac{J_{sc}}{J_0} + 1) \tag{7}$$

Here K, T, and q represent the Boltzmann constant, temperature, and electron charge, respectively. The rest of the parameters, *JSC* and *Jo,* stand for the short circuit current and reverse saturation current.

### 3.7. J-V and QE characteristics

The performance can be easily evaluated by the J-V curve of any solar structure. By analyzing the J-V curve, performance parameters, efficiency optimization, and loss mechanism understanding can be determined [15], [51]. This tool is useful for the validation of theoretical and practical cell performance. Fig.9(a) shows the optimized four structures J-V curve. All four structures show the same kind of characteristics and performance, which is evident from the Fig.9(a). Due to the less influence of buffer material on PCE, a similar type of J-V curve may be found.

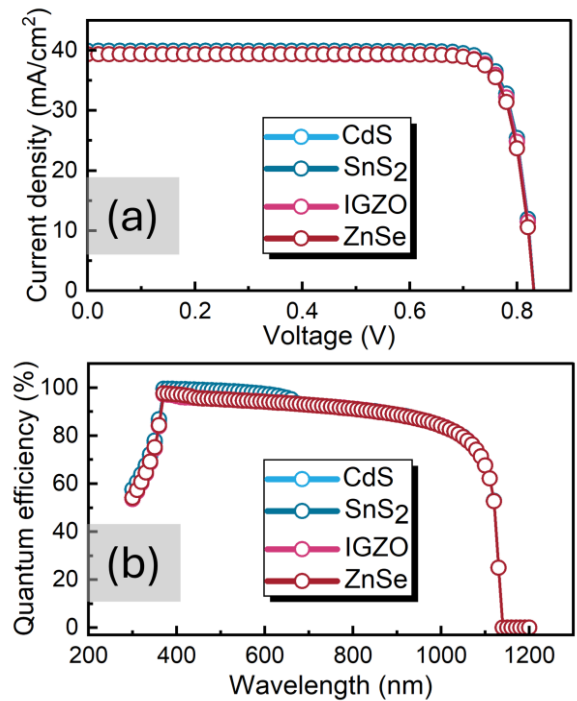

**Fig. 9.** (a) Current-voltage (J-V) characteristics and (b) quantum efficiency (QE) curves for the four proposed solar cell configurations, illustrating their electrical performance and light absorption.

Quantum efficiency (QE) is another benchmark for understanding the photon absorption quantity. QE refers to the absorption of photons to the incident of photons from the solar spectrum. Fig.9(b) shows the QE curve for the simulated structures. This shows a very excellent performance almost for all structures. Comparing them, $SnS_2$ shows an excellent QE between the light wavelength 370nm to 660nm with an amount of absorption beyond 95% of total incident photons. Configuration with CdS buffer material is also remarked as the second-best photon absorber among the structures. Above 1130 nm, the absorption for all structures is almost terminated. In summary, every optimized cell structure has almost identical characteristics and performance in this study.

### 3.8. Generation and recombination rate

Several characteristics of a solar cell depend on the rate of generation and recombination of the charge carriers. Fig.10(a-b) shows the generation and recombination profile with respect to position for the optimized structure of all four structures. The excitation of electrons from the valance band (VB) to the conduction band (CB) and leaving a hole in VB is the key process for the generation profile. In comparison, the highest amount of generation has been denoted near 1.57µm to 1.6µm for $SnS_2$. The rest of the materials follow a similar kind of trend, and the highest amount of absorption can be found near 1.5µm to 1.7µm. Since the largest number of electrons are produced at that particular point due to the enhanced capacity to absorb photons, the rate of generation reaches the peak in that spectrum. Consequently, an increased generation rate due to efficient light trapping results in an increase of short circuit current density (JSC) and, therefore, increased power output of the cell. However, recombination is when the electron-hole pairs (EHP) recombine before they can contribute to the circuit current. Eventually, it reduces the chances of having free charges, hence affecting both $V_{OC}$ and FF. Higher recombination rates due to defects or poor substrate quality material are inefficient since they cause loss of energy and decrease the overall PCE of the cell. To improve performance further, solar cells need to maximize carrier generation through superior light trapping and minimize recombination by excellent material quality, defect passivation, and better cell architecture. For the generation profile G(x) creation, SCPAS used the following equation

$$G(\lambda, x) = \alpha(\lambda, x).N_{phot}(\lambda, x) \tag{8}$$

Where $\alpha(\lambda, x)$ is the absorption profile and $N_{phot}(\lambda, x)$ is the photon flux from the incident sunlight.

Fig.10(b) shows the recombination characteristics for all simulated structures. The recombination process refers to the generated electrons and holes recombined and vanishing from CB to VB. The recombination process is the main culprit in reducing the carrier lifetime.

For the range between 1.6μm to 1.72μm, all peak recombination occurs for all structures. The early depth peak recombination can be exhibited by CdS at 1.6μm with an amount of $2.5\times10^{20}$ $cm^{-3}s^{-1}$. The other two configurations with IGZO and ZnSe buffer materials have also shown the highest amount of recombination at the same depth of position. Lastly, $SnS_2$ performs the peak recombination at 1.75μm depth. Increased recombination at the back contact of CZTSSe solar cells, around 1.6 μm deep, results from factors including grain boundary defects, inadequate photon absorption, and incorrect band alignment [25], [52], [53]. These problems diminish efficiency by increasing carrier losses. Proposed solutions include back contact passivation using materials such as $Al_2O_3$, improved absorber crystallinity via optimized annealing, and band alignment enhancement employing intermediate layers like ZnO or $TiB_2$[54]. Collectively, addressing these variables may significantly enhance the performance of CZTSSe solar cells.

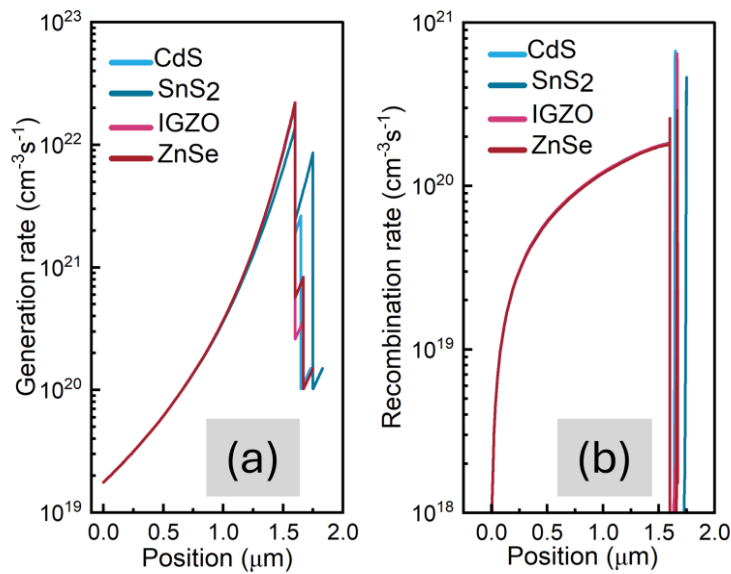

**Fig. 10.** Generation and recombination profiles for different solar cell configurations, highlighting the carrier generation rates and recombination processes within the device structures.

### 3.9. Comparison with prior studies

The previous structure of CZTS-based solar cells has attained a significant amount of practical and theoretical efficiency both in experimental and simulation work. Some of the cells show an excellent PCE with a higher amount of $V_{OC}$, FF, and $J_{SC}$. Table 2 represents an overview of recent and previous works. In the group of the previous study by W. Wang *et al.* in [55] and Y. Gong *et al.* in [56] experimental studies have shown a PCE of 12.6% and 12.5% with conventional CdS buffer layer for both CZTSSe and ACZTSSe absorber layer respectively. The previous simulation result shows a superb and excellent PCE. A. Ait Abdelkadir *et al*. [57] recorded a booming efficiency of 32.98%. This efficiency can be exhibited because of the addition of the back surface field (BSF) layer of NiO. BSF layer passivates recombination sites at interfaces, so decreasing carrier losses and enhancing current flow, while offering optimal band alignment for effective charge transfer. By performing as an intermediary layer, it eliminates interface defects and preserves the absorber layer from direct contact with the back contact.

**Table. 2.** Performance summary of CZTSSe-based solar cells from previous studies and this work, including absorber and buffer materials, (V), (mA/cm²), FF (%), PCE (%), study type, and references.

| Absorber Material | Buffer Material | $V_{OC}$ (V) | $J_{SC}$ (mA/cm²) | FF (%) | PCE (%) | Type | References |
|---|---|---|---|---|---|---|---|
| ACZTSSe | CdS | 0.54 | 32.1 | 72.3 | 12.5 | Experimental | [56] |
| CZTSe | CdS | 0.433 | 40.6 | 67.3 | 11.95 | Experimental | [60] |
| CZTSSe | CdS | 0.513 | 35.2 | 69.8 | 12.6 | Experimental | [55] |
| CZTSSe | ZnSnO | 0.455 | 36.4 | 69.3 | 11.2 | Experimental | [61] |
| CZTS | ZnCdS | 0.683 | 22.2 | 66.3 | 10.1 | Experimental | [62] |
| CZTSSe | CdS | 0.731 | 32.1 | 74.4 | 17.5 | Simulation | [63] |
| CZTS | $MoS_2$ | 0.85 | 25.30 | 84.76 | 18.27 | Simulation | [10] |
| CZTSe | CdS | 0.83 | 45.41 | 86.49 | 32.98 | Simulation | [57] |
| CZTSSe | $TiO_2$ | 0.72 | 42.61 | 74.93 | 23.13 | Simulation | [38] |
| CZTSSe | ZnSe | 0.72 | 42.58 | 72.68 | 22.42 | Simulation | [38] |
| $C_2N$ | IGZO | 1.22 | 17.65 | 84.43 | 18.22 | Simulation | [64] |
| CZTS | $TiO_2$ | 1.099 | 31.89 | 87.85 | 30.79 | Simulation | [58] |
| CZTSSe | $In_2S_3$ | 0.807 | 35.92 | 84.17 | 24.42 | Simulation | [14] |
| CZTSSe | ZnSe | 0.804 | 35.89 | 81.58 | 23.55 | Simulation | [14] |
| CZTSSe | CdS | 0.831 | 39.44 | 85.35 | 28.0 | Simulation | This work |
| CZTSSe | $SnS_2$ | 0.832 | 39.93 | 85.4 | 28.38 | Simulation | This work |
| CZTSSe | IGZO | 0.831 | 39.35 | 85.34 | 27.94 | Simulation | This work |
| CZTSSe | ZnSe | 0.831 | 39.36 | 84.94 | 27.82 | Simulation | This work |

Furthermore, it may enhance light control by performing as an optical spacer or reflector, so boosting photon absorption in the active zone. Above 30% is also reported in Ref. [58] with SnS as the back surface passivation layer. In our study, the CZTSSe absorber with $SnS_2$ buffer layer has shown the highest amount of PCE of 28.38%. All similar kind of nearby performance is shown because the main absorber remains constant for all structures, and the buffer layer cannot contribute that much like the absorber layer. One of the major targets of this study is to find a potential replacement for conventional CdS buffer material. The configuration of i-ZnO/SnS$_2$/CZTSSe/back contact can be a potential option to replace the cadmium-free CdS buffer layer-based solar cell. Additionally, SnS$_2$ offers a comparatively low manufacturing cost and a non-toxic alternative for chemical vapor deposition (CVD), chemical bath deposition, and spray pyrolysis deposition techniques in practical cell development [59].

4. **Conclusion**

This study utilized the SCAPS-1D simulator to analyze the performance of CZTSSe-based solar cells, with CZTSSe as the absorber material featuring a bandgap of 1.096 eV. Four distinct buffer materials—CdS, SnS$_2$, IGZO, and ZnSe—were evaluated based on energy band alignment. The analysis considered the effects of absorber defect density, buffer thickness, donor density, and defect density, concluding that these factors have minimal impact on solar cell efficiency. The optimal absorber thickness was determined to be 1600 nm for all configurations, while the ideal acceptor density of the absorber was identified to achieve maximum efficiency. After comprehensive optimization, the i-ZnO/SnS$_2$/CZTSSe/Au configuration demonstrated the highest PCE of 28.38%. Notably, the i-ZnO/CdS/CZTSSe/Au, i-ZnO/IGZO/CZTSSe/Au, and i-ZnO/ZnSe/CZTSSe/Au structures also achieved significant efficiencies of 28.00%, 27.94%, and 27.82%, respectively. Stability analyses revealed excellent performance across variations in series and shunt resistance, with optimal efficiency observed at ambient temperature. This detailed investigation provides valuable insights into the development of cost-effective and high-performance CZTSSe-based solar cells, paving the way for advancements in efficient and sustainable photovoltaic technologies.

**Acknowledgment**

The authors would like to acknowledge the use of SCAPS-1D as the primary tool for numerical modelling and simulation in this study. SCAPS-1D, a one-dimensional solar cell modelling software, was developed by Prof. Marc Burgelman and his team at the Department of Electronics and Information Systems (ELIS), University of Ghent, Belgium.